# Magic Fountain

**Roman Ya. Kezerashvili**
*New York City Technical College, The City University of New York*
*and*
**Alexander Sapozhnikov**
*Brooklyn College, The City University of New York*

At last North East Regional Conference of the American Association of Physics Teachers at Princeton University, at the fall meeting of the New England Section of the APS and AAPT at Milton Academy and at the 123rd National Meeting of the American Association of Physics Teachers was presented a simple exited experiment of magic Hero's fountain [1]. This demonstration took and attracted attention of the audience and many questions were raised about it construction and action. We have been asked to white comments about that demonstration. All these motivated us to write this note.

Hero of Alexandria (his name is also spelled Heron) was a Greek mathematician and scientist. He is best known today for an important theorem in plane geometry, also known as Heron's Formula (Hero's Formula), for finding the area of a triangle in terms of its sides. Hero has done his work in Alexandria, Egypt, and has written at least 13 works on mathematics, mechanics, and physics. He developed various mechanical devices, including the aelopile, a rotary steam engine, and water jet produced by mechanically compressed air, -apparatus known as Hero's fountain. He dealt with a number of such engineering devices in his book *Pneumatica* [2]. Hero's fountain, a pneumatic apparatus in which a vertical jet of water is produced and sustained by air pressure, and air forces water above its source.

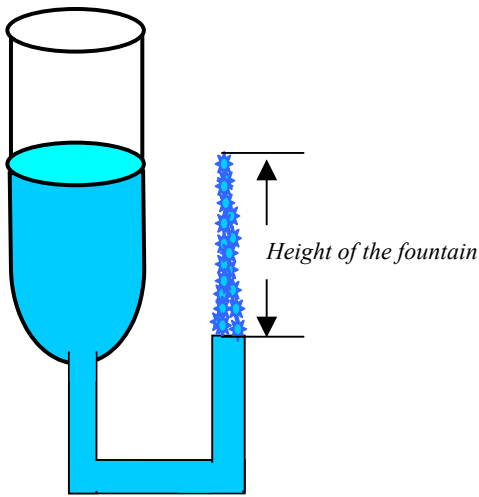

*Height of the fountain*

Fig. 1. Gravity-fed Fountain

Well known that the pressure of static fluid is the same at any horizontal level within connected vessels. This principle is underlying of a gravity-fed fountain shown in Fig.1. When the water is projected vertically from the pipe connected to open reservoir it reaches the maximum height that equal to an altitude of the top of water level in the open reservoir. The height of the fountain is a distance between the nozzle of the fountain and horizontal level of the open reservoir. If a fountain is not gravity-fed you need a pump to produce a difference of pressure to supply the fountain.

Today you can find a numerous publications and on the Internet (for example Refs. 3, 4, 5, 6, 7), which are describing operation and different modified constructions of Hero's fountain. The fountain invented by the ancient Greek mechanist doesn't have a vessel above a fountain nozzle or a pump and can be considered as one of the best demonstrations of the properties of liquids, explicitly of Pascal's and Bernoulli's principles, the topic that always discussed in introductory college and university physics. See, for example, Refs. 8 and 9. The construction of our Hero's fountain is rather different than the ancient original one [2] or a traditional demonstration [10] though the principle of the action is the same.

The principle of its action and explanation could be found in the construction of this fountain. Figure 2 incorporates a scheme for the Hero's fountain. The fountain consists of three parts: a cup *A* with fountain tubes, and two vessels *B* and *C* with partially filled by water. The parts are connected with two flexible hoses as shown in Fig.2: the cup *A* with lower vessel *C* and vessel *C* with upper vessel *B*. The vessel *B* with cup can be placed on a table and the other one below the level of the table. The cup *A* is maintained on the upper vessel *B* and connected with the lower vessel *C* by a flexible hose. Initially the pressure in both vessels *B* and *C* equal to atmospheric pressure. When you fill up the cup *A* with water the water from the cup *A* flows down to the lower vessel *C*, which contains air, and produces the additional hydrostatic pressure $P_2 = \rho g h_2$, where . is the density of water. According the Pascal's principle this additional pressure is transmitted undiminished to all directions and therefore to the air inside of the vessel *C*. As a result this pressure forces a confined air up from vessel C to the upper vessel *B*. The forced air from the lower vessel *C* squeezes the air in the upper vessel *B*, and forces the water spout out of the fountain upper tube. At this moment the hydrostatic pressure in the upper vessel *B* is equal $P_1 = \rho g h_1$. Thus, the pressure of the water in the fountain is the difference of the hydrostatic pressures in the vessels *C* and *B*. Therefore

$$\Delta P = P_2 - P_1 = \rho g h_2 - \rho g h_1 = \rho g (h_2 - h_1). \qquad (1)$$

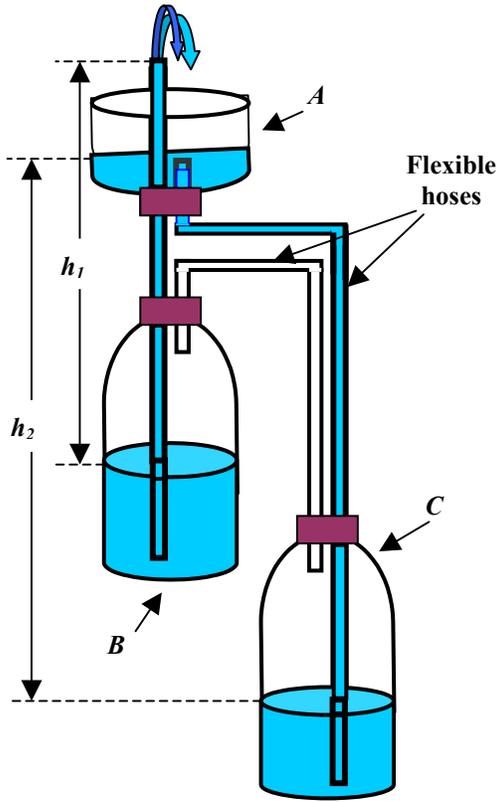

Fig. 2. Schematic presentation for the Hero's Fountain

In other words the pressure $\Delta P$ compresses the air in the upper vessel *B* and drives the fountain. If we neglect the height of the nozzle of the fountain above the water level in the cup *A*, difference of heights $h_2 - h_1$ will be equal to the altitude of the top of the water level in the vessel *A* measured with respect to the water level in the vessel *C*. By changing this altitude we can maintain the fountain.

Hero's fountain is also good demonstration for the Bernoulli's Principle. The Bernoulli's Principle is a result of application of the work-energy theorem to a unit volume of a moving fluid. It simply states that the work done on a unit volume of fluid by the surrounding fluid is equal to the sum of the changes in kinetic and potential energies per unit volume that occur during the flow. Let us apply Bernoulli's Principle for the water in reservoir *A* and a stream of water from the nozzle. The potential energy of a unit volume of water on the level of the reservoir *A* and the nozzle are respectively $\rho g h_2$ and $\rho g h_1$, and kinetic energies are respectively zero and $\dfrac{\rho v^2}{2}$.

Thus we have from Bernoulli's equation

$$P_{atm} + \frac{\rho v^2}{2} + \rho g h_1 = P_{atm} + \rho g h_2 , \qquad (2)$$

where $P_{atm}$ is atmospheric pressure. The speed of the stream of water from the nozzle of the fountain tube can be found easily from equations (2) and (1) as

$$v = \sqrt{2g(h_2 - h_1)} = \sqrt{\frac{2\Delta P}{\rho}} . \qquad (3)$$

Deriving equation (3) we assuming that a confined air is uncompressible and we neglecting the friction. Of course, the real velocity of water from the nozzle will be less than it is given by equation (3). Consider again the fluid as an ideal by moving the vessel $C$ up and down we can change the gauge pressure $\Delta P$ and as a result the speed of the water from the nozzle will change. When pressure $\Delta P$ increases the speed of water increases and the water rises higher with respect to the nozzle creating the fountain effect and vice versus. If bottle $C$ is lowered even further, then the hydrostatic pressure increases inside $C$, and consequently inside $B$ as well, causing the fountain to spray even higher. Conversely, if bottle $C$ is raised relative to the bottle $B$, the hydrostatic pressure decreases, and the fountain spray diminishes.

We can visualize the Hero's fountain in terms of potential and kinetic energy. At the beginning the water is at its lowest potential energy in the lower vessel $C$. The water in the upper vessel $B$ is stationary. It does not begin flowing until we activated the fountain by adding additional potential energy to the system when we pouring water into the cup $A$. This is the activation energy of the system. Once the cup $A$ and tubing is full, the potential energy between the upper and lower water levels in the vessels B and C determines velocity of droplets moving upwards. At the nozzle the kinetic energy of the water is largest. The potential energy of the water is largest at the highest point in the stream of the fountain of water**.**

To make the magic fountain need three 2-liter plastic soda bottles, three rubber stoppers with two 1/8" holes each, two plastic tubing about 2.5 ft long and 1/8" glass tube about 3 ft long. Glass tube will be cut on five pieces: 16", 13", 3" and two 3.1/2". From the 3.1/2" glass tube pieces make right angle bends with the side length 1.1/2". The 16" glass tube may be replace by metal tube of the same diameter. Insert two right angle bends into each of two rubber stoppers and then insert 16" fountain tube into second hole of these stoppers. The cup $A$ of the fountain is made from the upper 1/3 part of the plastic soda bottle by cutting it. Setting the cup $A$ upright, connect it and partially filled plastic soda bottle $B$ mouth to mouth using two rubber stoppers with right angle bends and fountain tube. Insert 13" and 3" glass tubes into the third rubber stopper and place it into the empty plastic coda bottle $C$. Finally connect the glass tubes of the bottle $C$ and glass tube of bottle $B$ and cup $A$ with the plastic tubes. To start the fountain you should pour the glass of water into the cup $A$ and stand the empty bottle $C$ lower then one with water. The fountain works when leaks of water or air are escape. The magic fountain will act since all water of the upper bottle flows into the empty bottle. To restart the fountain you can replace the bottles and again pour water into the cup $A$. The fountain works better and longer if the nozzle of the fountain tube is narrow. It spouts into the open air for a few minutes. For more impression use colored water or Coca-Cola. This dramatic presentation your whole class will love.